\documentclass[preprint,12pt,review,authoryear]{elsarticle}

\usepackage{amssymb}
\usepackage{graphicx}
\usepackage{epstopdf}
\usepackage{float}
\usepackage{caption}
\usepackage{subcaption}

\journal{International Journal of Solids and Structures}

\makeatletter
\def\@author#1{\g@addto@macro\elsauthors{\normalsize%
    \def\baselinestretch{1}%
    \upshape\authorsep#1\unskip\textsuperscript{%
      \ifx\@fnmark\@empty\else\unskip\sep\@fnmark\let\sep=,\fi
      \ifx\@corref\@empty\else\unskip\sep\@corref\let\sep=,\fi
      }%
    \def\authorsep{\unskip,\space}%
    \global\let\@fnmark\@empty
    \global\let\@corref\@empty  %% Added
    \global\let\sep\@empty}%
    \@eadauthor={#1}
}
\makeatother

\begin{document}

\newcommand{\hilight}[1]{\colorbox{yellow}{#1}} %To highlight references in the text

\begin{frontmatter}



\title{Modeling damage and fracture within strain-gradient plasticity}

% Another option "Modeling damage and fracture accounting for the plastic size effect"


\author{E. Mart\'{\i}nez-Pa\~neda\corref{cor1}}
\ead{martinezemilio@uniovi.es}

\author{C. Beteg\'on}

\address{Department of Construction and Manufacturing Engineering, University of Oviedo, Gij\'on 33203, Spain}

\cortext[cor1]{Corresponding author. Tel: +34 985 18 19 67; fax: +34 985 18 24 33.}

\begin{abstract}
In this work, the influence of the plastic size effect on the fracture process of metallic materials is numerically analyzed using the strain-gradient plasticity (SGP) theory established from the Taylor dislocation model. Since large deformations generally occur in the vicinity of a crack, the numerical framework of the chosen SGP theory is developed for allowing large strains and rotations. The material model is implemented in a commercial finite element (FE) code by a user subroutine, and crack-tip fields are evaluated thoroughly for both infinitesimal and finite deformation theories by a boundary-layer formulation. An extensive parametric study is conducted and differences in the stress distributions ahead of the crack tip, as compared with conventional plasticity, are quantified. As a consequence of the strain-gradient contribution to the work hardening of the material, FE results show a significant increase in the magnitude and the extent of the differences between the stress fields of SGP and conventional plasticity theories when finite strains are considered. Since the distance from the crack tip at which the strain gradient significantly alters the stress field could be one order of magnitude higher when large strains are considered, results reveal that the plastic size effect could have important implications in the modelization of several damage mechanisms where its influence has not yet been considered in the literature. 
\end{abstract}


\end{frontmatter}


\section{Introduction}
\label{Introduction}

Experiments and direct-dislocation simulations have demonstrate that metallic materials display a strong size effect at the micro- and sub-micron scales. Attributed to geometrically necessary dislocations (GNDs) associated with non-uniform plastic deformation; this size effect is especially significant in fracture problems since the plastic zone adjacent to the crack tip is physically small and contains strong spatial gradients of deformation. Since conventional plasticity possesses no intrinsic material length, several continuum strain-gradient plasticity (SGP) theories have been developed through the years to incorporate some length-scale parameters in the constitutive equations. Most of them can be classified as a function of their approach as phenomenological \citep{FH93,FH97,FH01} or mechanism-based \citep{G99,H00}.

The experimental observation of cleavage fractures in the presence of significant plastic flow \citep{E94,K02} has aroused significant interest in the influence of the plastic strain gradient on crack-tip stresses \citep{WH97,C99,W01}. \citet{J01} investigated the crack-tip field by a mechanism-based strain-gradient (MSG) plasticity theory established from the Taylor dislocation model. Their investigation showed that GNDs near the crack tip promoted strain-hardening and that the GNDs led to a much higher stress level in the vicinity of the crack than that predicted by classical plasticity. \citet{Q04} implemented the lower-order conventional theory of mechanism-based strain-gradient (CMSG) plasticity \citep{H04} that does not involve higher-order stresses and where the plastic strain gradient is involved through the incremental plastic modulus. They showed that the higher-order boundary conditions have essentially no effect on the stress distribution at a distance greater than 10 nm from the crack tip, well below the lower limit of physical validity of the SGP theories based on Taylor's dislocation model \citep{S01}. 

However, the aforementioned studies were conducted in the framework of the infinitesimal deformation theory, and although large deformations occur in the vicinity of the crack, little work has been done to investigate crack-tip fields under SGP to account for finite strains. \citet{H03} developed a finite deformation theory of MSG plasticity, but they were unable to reach strain levels higher than 10\% near the crack tip due to convergence problems. \citet{MG09} determined the range of material length scales where a full strain-gradient-dependent plasticity simulation is necessary in the finite strain version \citep{NR04} of the SGP theory of \citet{FH01}. \citet{PY11} used the element-free Galerkin method to analyze the crack-tip stresses through a lower-order gradient plasticity (LGP) model \citep{YC00} and they showed that the known elastic-plastic fracture mechanics parameter $G$ can be directly applied to the crack assessment under strain-gradient plasticity for both infinitesimal and finite deformation theories. 

Moreover, identifying and quantifying the relation between material parameters and the physical length over which gradient effects prominently enhance crack-tip stresses is essential in rating their influences on crack-growth mechanisms, and for rationally applying SGP theories to predict damage and fracture. This has been done recently by \citet{K08} for the phenomenological SGP theory of \citet{FH01} within the small-strain theory. But, as the strain gradient increases the resistance to plastic deformation thereby lowering crack-tip blunting, and consequently avoiding the local stress triaxiality reduction characteristic of the conventional plasticity predictions, it is imperative to quantify the distance ahead of the crack tip where the plastic size effect significantly alters the stress distribution accounting for finite strains.

In this work, the influence of the plastic strain gradient on the fracturing process of metallic materials is numerically analyzed in the framework of small- and large-deformations by the CMSG theory. An extensive parametric study is conducted and differences in the stress distributions ahead of the crack tip, compared with conventional plasticity, are quantified. Implications of the results on fracture- and damage-modeling are thoroughly discussed. 

\section{Conventional theory of mechanism-based strain gradient}
\label{CMSG theory description}

The conventional theory of mechanism-based strain-gradient plasticity (CMSG) is based on the Taylor dislocation model but does not involve higher-order stresses. Therefore, the plastic strain gradient appears only in the constitutive model and the equilibrium equations and boundary conditions are the same as the conventional continuum theories \citep{H04}. 

The dislocation model of \citet{T38} gives the shear-flow stress $\tau$ in terms of the dislocation density $\rho$ as:

\begin{equation}
\tau = \alpha \mu b \sqrt{\rho}
\end{equation}

where $\mu$ is the shear modulus, $b$ is the magnitude of the Burgers vector, and $\alpha$ is an empirical coefficient that takes values between 0.3 and 0.5. The dislocation density comprises the sum of the density $\rho_S$ for statistically stored dislocations and the density $\rho_G$ for geometrically necessary dislocations:

\begin{equation}
\rho = \rho_S + \rho_G,
\end{equation}

with $\rho_G$ related to the effective plastic strain gradient $\eta^{p}$ by: 

\begin{equation}
\rho_G = \overline{r}\frac{\eta^{p}}{b}
\end{equation}

where $\overline{r}$ is the Nye-factor that is assumed to be approximately 1.90 for face-centered-cubic (fcc) polycrystals.

The tensile flow stress $\sigma_{flow}$ is related to the shear-flow stress $\tau$ by:

\begin{equation}
\sigma_{flow} =M\tau,
\end{equation}

$M$ being the Taylor factor, that equals 3.06 for fcc metals. Rearranging Eqs. (1-4) yields:

\begin{equation}
\sigma_{flow} =M\alpha \mu b \sqrt{\rho_{S}+\overline{r}\frac{\eta^{p}}{b}}.
\end{equation}

$\rho_{S}$ can be determined from (5), knowing the relation in uniaxial tension (where $\eta^{p}=0$) between the flow stress and the material stress-strain curve, as 

\begin{equation}
\rho_{S} = [\sigma_{ref}f(\varepsilon^{p})/(M\alpha \mu b)]^2,
\end{equation}

where $\sigma_{ref}$ is a reference stress and $f$ is a non-dimensional function of plastic strain $\varepsilon^{p}$ determined from the uniaxial stress-strain curve. Substituting in (5), $\sigma_{flow}$ yields:

\begin{equation}
\sigma_{flow} =\sigma_{ref} \sqrt{f^2(\varepsilon^{p})+l\eta^{p}}
\end{equation}

where $l$ is the intrinsic material length that provides a combination of the effects of elasticity ($\mu$), plasticity ($\sigma_{ref}$), and atomic spacing ($b$) and is given by:

\begin{equation}
l=M^2\overline{r}\alpha^2 \left(\frac{\mu}{\sigma_{ref}}\right)^2b=18\alpha^2\left(\frac{\mu}{\sigma_{ref}}\right)^2b.
\end{equation}

According to \citet{G99}, the effective plastic strain gradient $\eta^{p}$ is given by:

\begin{equation}
\eta^{p}=\sqrt{\frac{1}{4}\eta^{p}_{ijk} \eta^{p}_{ijk}}
\end{equation}

where the third-order tensor $\eta^{p}_{ijk}$ is obtained by:

\begin{equation}
\eta^{p}_{ijk}=\varepsilon^{p}_{ik,j}+\varepsilon^{p}_{jk,i}-\varepsilon^{p}_{ij,k}
\end{equation}

and the tensor for plastic strain equals:

\begin{equation}
\varepsilon^{p}_{ij}=\int \dot{\varepsilon}^{p}_{ij} dt.
\end{equation}

To avoid higher-order stresses, \citet{H04} used a viscoplastic formulation that gives the plastic strain rate $\dot{\varepsilon}^{p}$ in terms of the effective stress $\sigma_e$ rather than its rate $\dot{\sigma}_e$. Also, to remove the strain-rate- and time-dependence, a viscoplastic-limit is used by replacing the reference strain with the effective strain rate $\dot{\varepsilon}$:

\begin{equation}
\dot{\varepsilon}^{p} = \dot{\varepsilon} \left [\frac{\sigma_e}{\sigma_{ref} \sqrt{f^{2}(\varepsilon^{p})+l\eta^{p}}} \right]^{m}
\end{equation}

This procedure is merely for mathematical convenience and differences are negligible for a large value of the rate-sensitivity exponent ($m\geq20$). Considering that the volumetric- and deviatoric- strain rates are related to the stress rate in the same way as in classical plasticity, the constitutive equation of the CMSG theory yields:

\begin{equation}
\dot{\sigma}_{ij}=K\dot{\varepsilon}_{kk}\delta_{ij}+2\mu \left\{\dot{\varepsilon}'_{ij} - \frac{3\dot{\varepsilon}}{2\sigma_e}\left[\frac{\sigma_e}{\sigma_{flow}} \right]^{m}\dot{\sigma}'_{ij} \right\}
\end{equation}

As it is based on the Taylor dislocation model, which represents an average of dislocation activities, the CMSG theory is only applicable at a scale much larger than the average dislocation spacing. For common values of dislocation density in metals, the lower limit of physical validity of the SGP theories based on Taylor's dislocation model is approximately 100 nm. 

\section{Crack-tip fields with infinitesimal strains}
\label{Computational results for small strains}

Crack-tip fields are evaluated in the framework of the finite element method by a boundary-layer formulation, where the crack region is contained by a circular zone and a mode-I load is applied at the remote circular boundary through a prescribed displacement: 

\begin{equation}
u(r,\theta)=K_I \frac{1+\nu}{E} \sqrt{\frac{r}{2\pi}}cos\left(\frac{\theta}{2}\right)(3-4\nu-cos\theta)
\end{equation}

\begin{equation}
v(r,\theta)=K_I \frac{1+\nu}{E} \sqrt{\frac{r}{2\pi}}sin\left(\frac{\theta}{2}\right)(3-4\nu-cos\theta)
\end{equation}

$u$ and $v$ being the horizontal and vertical components of the displacement boundary condition, respectively; $r$ and $\theta$ the radial and angular coordinates of a polar coordinate system centered at the crack tip, $E$ and $\nu$ the elastic properties of the material, and $K_I$ the stress intensity factor that quantifies the remote applied load. 

The material model is implemented in the commercial finite element package ABAQUS via its user-material subroutine UMAT. Since higher-order boundary conditions are not involved, the governing equations of the CMSG theory are essentially the same as those in conventional plasticity. The plastic strain gradient is obtained by numerical differentiation within the element: the plastic strain increment is interpolated through its values at the Gauss integration points in the isoparametric space and afterwards the increment in the plastic strain gradient is calculated by differentiation of the shape function. Another possible implementation scheme lies in using $C^0$ finite elements incorporating the effect of the strain gradient as an extension of the classical FE formulation \citep{S06,S06b}.

Plane strain conditions are assumed and only the upper half of the circular domain is modeled because of symmetry. An outer radius of $R$=42 mm is defined and the entire specimen is discretized using 1580 eight-noded quadrilateral plane-strain elements with reduced integration (CPE8R). As seen in Fig. 1, to accurately characterize the strain-gradient effect, a very fine mesh is used near the crack tip, where the length of the smallest element is approximately 10 nm. 

To compare and validate our numerical implementation, we have employed the same material properties considered by \citet{Q04} in the present study. Thus, if the stress-strain relation in uniaxial tension can be written as:

\begin{equation}
\sigma=\sigma_{ref} f(\varepsilon^{p})=\sigma_Y \left( \frac{E}{\sigma_Y} \right)^{N} \left(\varepsilon^{p}+\frac{\sigma_Y}{E} \right)^{N}
\end{equation}

where $\sigma_Y$ is the initial yield stress and $N$ is the strain hardening exponent.  $\sigma_{ref}=\sigma_Y \left(E/\sigma_Y\right)^{N}$ is the assumed reference stress, and $f(\varepsilon^{p})=\left(\varepsilon^{p}+\left(\sigma_Y/E\right)\right)^{N}$; the material parameters being $\sigma_Y=0.2\%$ of $E$, $\nu=0.3$, $N=0.2$, $m=20$, $b=0.255$ nm, and $\alpha=0.5$, which give an intrinsic material length of $l=3.53$ $\mu$m according to (8). 

\begin{figure}[htbp]
\centering
\includegraphics{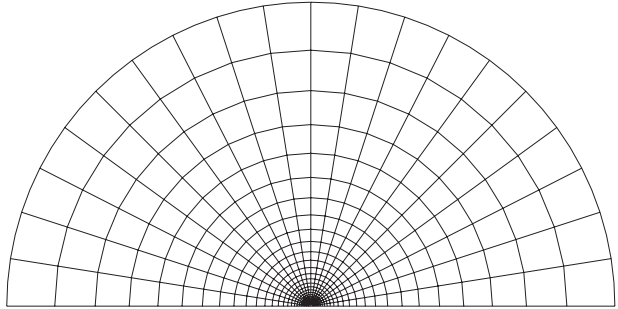}
\caption{Finite element mesh for the boundary layer formulation}
\label{fig:Fig1}
\end{figure}

Fig. 2 shows the hoop stress $\sigma_{\theta\theta}$ distribution ahead of the crack tip ($\theta=0^{\circ}$) under a remote load of $K_{I}=17.3\sigma_Y \sqrt{l}$ for both CMSG and classical plasticity theories; $\sigma_{\theta\theta}$ is normalized to the material yield stress and the distance to the crack tip $r$ ranges from 0.1 $\mu$m, the lower limit of CMSG plasticity, to 100 $\mu$m. As depicted in Fig. 2, the stress-field predicted by the CMSG theory agrees with the estimations of Hutchinson, Rice, and Rosengren (HRR) away from the crack tip, but becomes much larger within 1 $\mu$m distance from it. Indeed, the stress level in the CMSG theory at $r=0.1 \, \mu$m is equal to 12$\sigma_Y$, which is high enough to trigger cleavage fracture as discussed by \citet{Q04}. Results agree with those obtained by \citet{Q04} and \citet{J01} for the CMSG and MSG theories, respectively, proving that higher-order boundary conditions do not influence crack-tip fields within its physical domain and thus validating the present numerical implementation. Note that the crack-tip stress-elevation obtained by the mechanism-based theory quantitatively agrees with the predictions of the phenomenological approach, but with length parameters 4-5 times the corresponding quantity in the Fleck-Hutchinson theory: $l_{MSG}\approx (4-5) l_{SG}$ (see \citealp{WQ04}). 

\begin{figure}[htbp]
\centering
\includegraphics{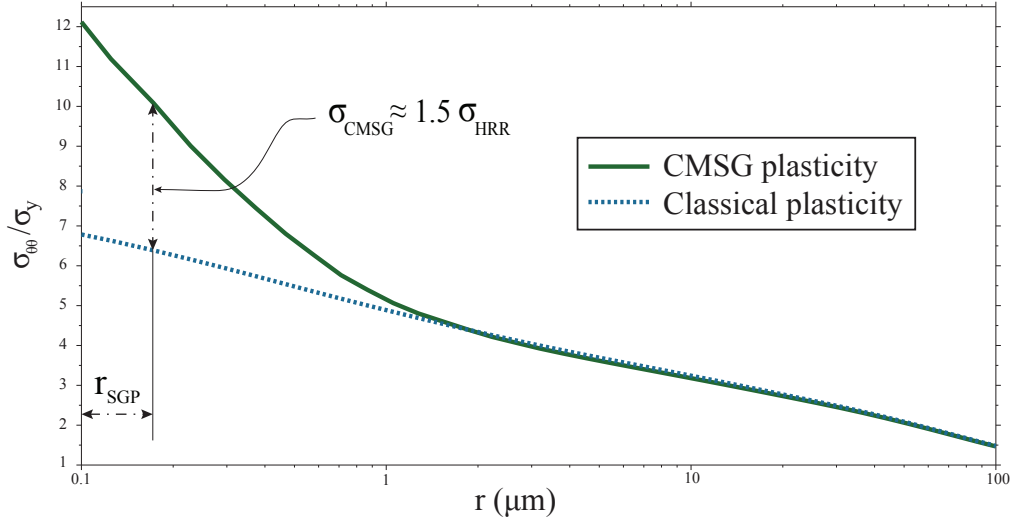}
\caption{$\sigma_{\theta\theta}$ distribution ahead of the crack tip for both CMSG and classical plasticity theories in small strains, $r$ being the distance to the crack tip in log scale for $K_{I}=17.3\sigma_Y \sqrt{l}$, $\sigma_Y=0.2\%$ of $E$, $\nu=0.3$, $N=0.2$ and $l=3.53$ $\mu$m}
\label{fig:Fig2}
\end{figure}

A parametric study covering several material properties, applied loads, and constraint conditions is conducted as a function of physical inputs to determine the influence of the strain gradient on crack-tip fields. As shown in Fig. 2, with the aim of quantifying the size of the region that is affected by the plastic size effect, the distance over which the stress is significantly higher than that predicted by conventional plasticity ($\sigma_{CMSG}> 1.5 \sigma_{HRR}$) is defined by $r_{SGP}$. Differences between the stress field obtained at a given point in the crack-tip region, for the CMSG theory ($\sigma_{CMSG}$), and the HRR field ($\sigma_{HRR}$) will depend on the following dimensionless terms:

\begin{equation}
\frac{\sigma_{CMSG}}{\sigma_{HRR}}=f \left(\frac{\sigma_Y}{E},N,\nu,\frac{l}{R},\frac{K_I}{\sigma_Y \sqrt{l}}\right)
\end{equation}

The material properties considered in Fig. 2 are taken as reference values and to avoid confusion as its corresponding variables are denoted with an asterisk. Also, to quantify the plastic size effect under different crack-tip constraint conditions, the stress-fields are evaluated through a modified boundary layer (MBL) formulation where the remote boundary is also dependent on the elastic $T$-stress \citep{BH91}:

\begin{equation}
u(r,\theta)=K_I \frac{1+\nu}{E} \sqrt{\frac{r}{2\pi}}cos\left(\frac{\theta}{2}\right)(3-4\nu-cos\theta)+T\left(\frac{1-\nu^2}{E}\right)R cos\theta
\end{equation}

\begin{equation}
v(r,\theta)=K_I \frac{1+\nu}{E} \sqrt{\frac{r}{2\pi}}sin\left(\frac{\theta}{2}\right)(3-4\nu-cos\theta)-T\left(\frac{\nu(1+\nu)}{E}\right)R sin\theta
\end{equation}

Fig. 3a shows the size of the domain influenced by the strain gradient plotted as a function of the applied load for the same configuration and material properties as above with $r_{SGP}$ normalized to the outer radius $R$ and the normalized applied stress intensity factor going from $K_{I}=30\sigma_Y^* \sqrt{l^*}$ to $K_{I}=300\sigma_Y^* \sqrt{l^*}$. The trend described by $r_{SGP}$ could be justified by the influence of geometrically necessary dislocations on plastic resistance. Since, as can be seen in (7) and (13), the plastic strain gradient $\eta^{p}$ is an internal variable of the constitutive equation of the CMSG theory which acts to increase the tangent modulus, hence reducing the plastic strain rate. Therefore, the plastic size effect translates into an additional hardening law, which causes an increase of the stress level that is enhanced as the applied load increases. Maintaining small-scale yielding (SSY) conditions, three load levels are considered in the analysis of subsequent parameters: $K_{I}=0.12\sigma_Y^* \sqrt{R}$, $K_{I}=0.6\sigma_Y^* \sqrt{R}$ and $K_{I}=1.2\sigma_Y^* \sqrt{R}$.

\begin{figure}
\makebox[\linewidth][c]{%
\begin{subfigure}[h]{0.55\textwidth}
                \centering
                \includegraphics{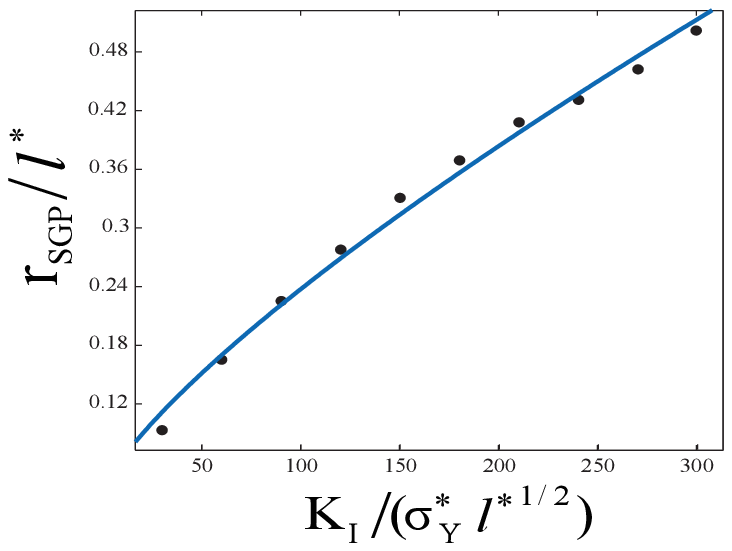}
                \caption{}
                \label{fig:Fig3a}
        \end{subfigure}
        \begin{subfigure}[h]{0.55\textwidth}
                \centering
                \includegraphics{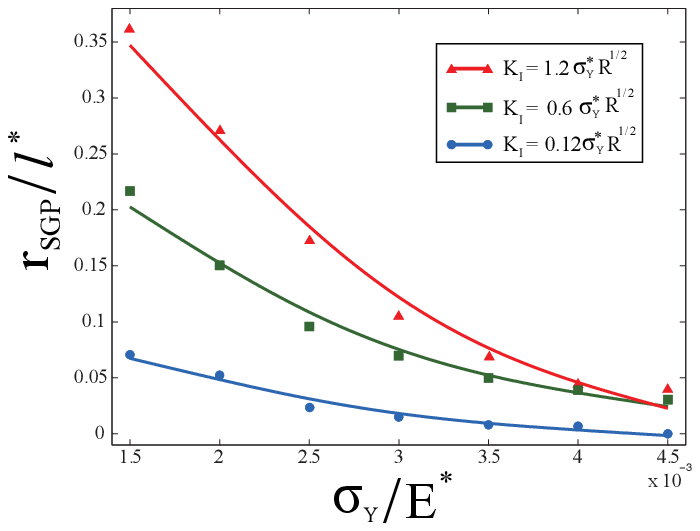}
                \caption{}
                \label{fig:Fig3b}
        \end{subfigure}
       }
\makebox[\linewidth][c]{%       
        \begin{subfigure}[h]{0.55\textwidth}
                \centering
                \includegraphics{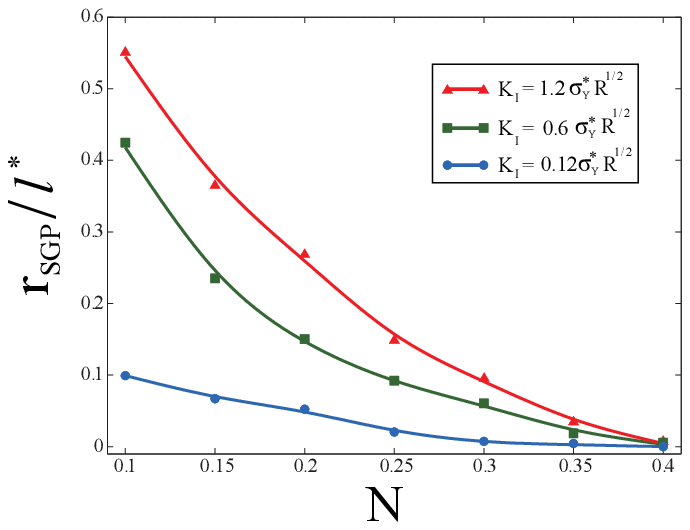}
                \caption{}
                \label{fig:Fig3c}
        \end{subfigure}
        \begin{subfigure}[h]{0.55\textwidth}
                \centering
                \includegraphics{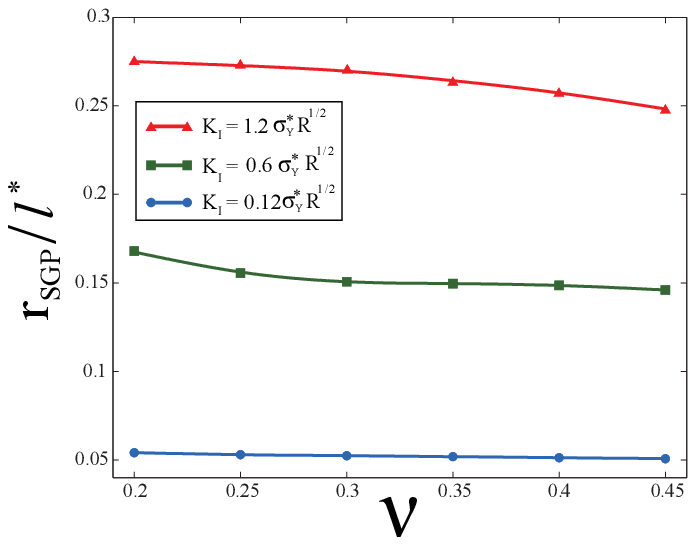}
                \caption{}
                \label{fig:Fig3d}
        \end{subfigure}
        }
\makebox[\linewidth][c]{%  
        \begin{subfigure}[h]{0.55\textwidth}
                \centering
                \includegraphics{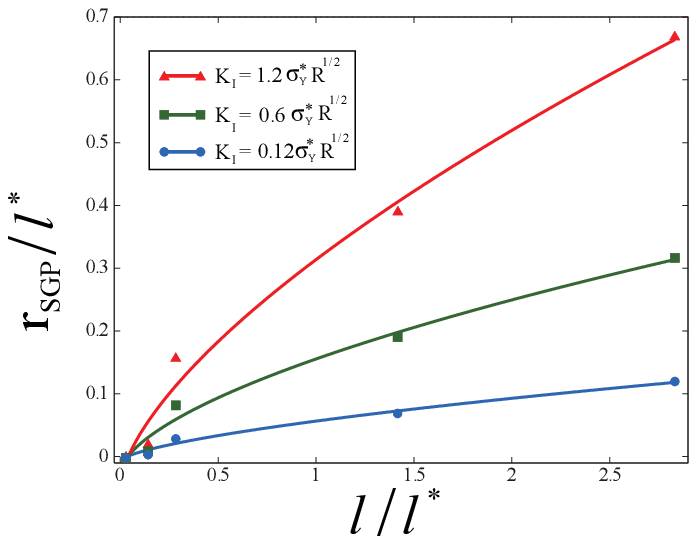}
                \caption{}
                \label{fig:Fig3e}
        \end{subfigure}
        \begin{subfigure}[h]{0.55\textwidth}
                \centering
                \includegraphics{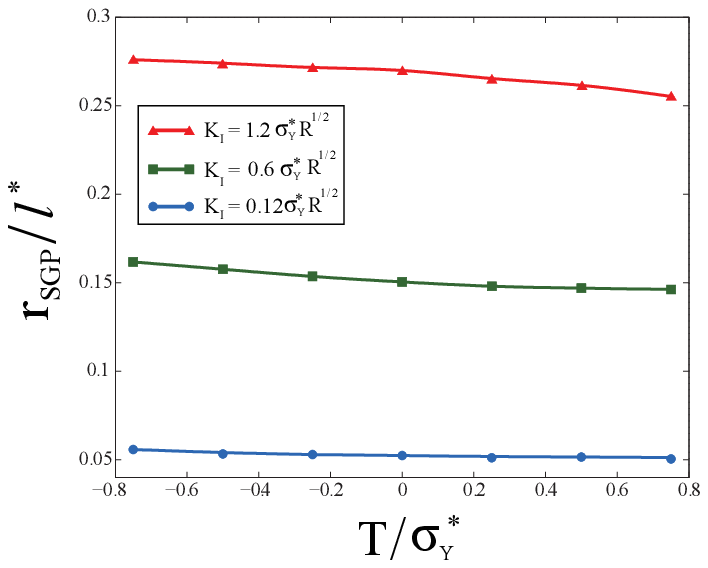}
                \caption{}
                \label{fig:Fig3f}
        \end{subfigure}
        }
        \caption{Distance ahead of the crack tip where the strain gradient significantly influences the stress distribution in small strains as a function of (a) applied load $K_I$, (b) yield stress $\sigma_Y$, (c) strain hardening exponent $N$, (d) Poisson's ratio $\nu$, (e) intrinsic material length $l$ and (f) $T$-stress.}\label{fig:Fig3}
\end{figure}

Fig. 3b shows the plots for the normalized relation between $r_{SGP}$ and the material elastic properties for different values of the yield stress $\sigma_Y$. The results show that as the value of the yield stress increases, the length of the domain where crack-tip fields are influenced by the size effect decreases. This is due to the fact that a higher value of $\sigma_Y$ causes a reduction in plastic deformation, hence downsizing the interval in which the strain gradient influences the tangent modulus. Note also that while an increase in the value of $\sigma_Y$ translates into a higher $\sigma_{flow}$ in conventional plasticity, the magnitude of the term accounting for the strain-gradient effect in (7) is independent of the material yield stress since the intrinsic material length $l$ (8) also depends on $\sigma_Y$. 

Fig. 3c illustrates the normalized distance over which the strain gradient significantly influences the stress distribution as a function of the strain hardening exponent $N$, with $N$ values varying between 0.1 and 0.4. As seen in Fig. 3c, the higher the work-hardening degree of the material the lower the extension of the influence of the plastic size effect on crack-tip fields. Since, as shown by \citet{S01} for the MSG theory and unlike the HRR field, the power of stress-singularity in CMSG plasticity is independent of $N$. This is because the strain gradient becomes more singular than the strain near the crack tip, and it dominates the contribution to the flow stress in (7), implying that the density of geometrically necessary dislocations $\rho_G$ around the crack tip is significantly larger than the density of statistically stored dislocations $\rho_S$.

Fig. 3d shows the variation of the normalized magnitude of the domain influenced by the size effect for different values of the Poisson's ratio $\nu$ (0.2-0.45). The results show that an increase in the Poisson's ratio leads to a reduction in the extension of the differences caused by the plastic size effect. This is a result of the Poisson's ratio influence on plastic deformation and its weight on the intrinsic material length (8). 

In Fig. 3e the normalized distance ahead of the crack tip, where the strain gradient influences the stress distribution, is plotted as a function of the intrinsic material length $l$. A range of values for $l$ of 0.1-100 $\mu$m is considered, since the scale at which the plastic size effect is observed is on the order of microns \citep{FH93}, and corresponds to the range of values that can take $l$ according to (8) for material properties common to metals. As expected, higher values of $l$ also result in higher values of $r_{SGP}$ since the influence of the term associated with the strain gradient inside the square root in (7) increases.

Fig. 3f shows the variation of the normalized size of the domain influenced by the strain gradient for different constraint situations. As can be seen, $r_{SGP}$ decreases as the constraint level increases because of the plastic-zone size dependence on the elastic $T$-stress \citep{W91}. However, the length of the domain where crack-tip fields are influenced by the size effect shows very low sensitivity to different crack-tip constraint conditions since changes on the $T$-stress value entail the same effect in both CMSG and HRR fields: negative $T$-stresses lead to a significant downward shift in the stress fields whereas positive values of $T$ slightly increase the stress level near the crack.

\section{Crack-tip fields with finite strains}
\label{Computational results for large strains}

Stress distributions in the vicinity of the crack are obtained in the framework of the finite deformation theory. Rigid body rotations for the strains and stresses are conducted by the \citet{HW80} algorithm and the strain gradient is obtained from the deformed configuration since the infinitesimal displacement assumption is no longer valid. 

The initial configuration and the background mesh of the boundary layer formulation are shown in Fig. 4. A very fine mesh of 6134 CPE8R elements is used to obtain accurate results. As seen in Fig. 5, the hoop stress $\sigma_{\theta\theta}$ distribution ahead of the crack line is obtained for both the CMSG and classical plasticity theories for the same material properties and loading conditions as in Fig. 2.

\begin{figure}[h]
        \centering
        \begin{subfigure}[h]{0.45\textwidth}
                \centering
                \includegraphics{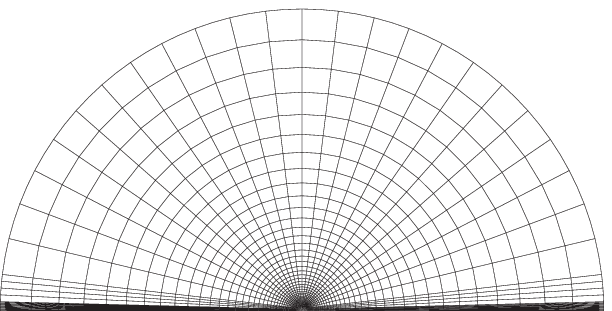}
                \caption{}
                \label{fig:Fig4a}
        \end{subfigure}
        \begin{subfigure}[h]{0.45\textwidth}
                \centering
                \includegraphics{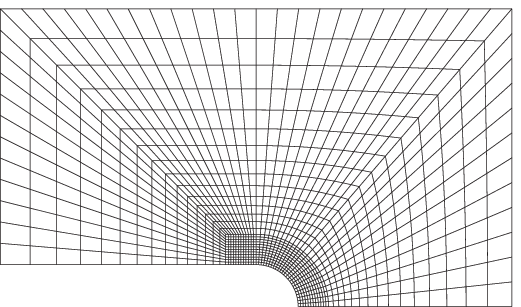}
                \caption{}
                \label{fig:Fig4b}
        \end{subfigure}
       
        \caption{Finite element mesh for the boundary layer formulation under large deformations: (a) complete model and (b) vicinity of the crack}\label{fig:Fig3}
\end{figure}

In classical plasticity \citep{M97}, large strains at the crack tip cause the crack to blunt, which reduces the stress triaxiality locally. However, because of the strain-gradient contribution to the work-hardening of the material, this behavior is not appreciated when the plastic size effect is considered. As proved by \citet{M97}, in conventional plasticity the crack-opening stress reaches a peak at approximately the same distance from the crack tip as the onset of the asymptotic behavior of the plastic-strain distribution. Therefore, as seen in Fig. 5, the strain gradient influences the stress distribution of the CMSG theory at approximately the same distance where a maximum of $\sigma_{\theta\theta}$ is obtained in conventional plasticity, significantly increasing the differences between the stress distributions of the SGP and classical plasticity theories; the magnitude of the distance where these differences occur: $r_{SGP}$ is one order of magnitude higher than that presented in Fig. 2.

\begin{figure}[htbp]
\centering
\includegraphics{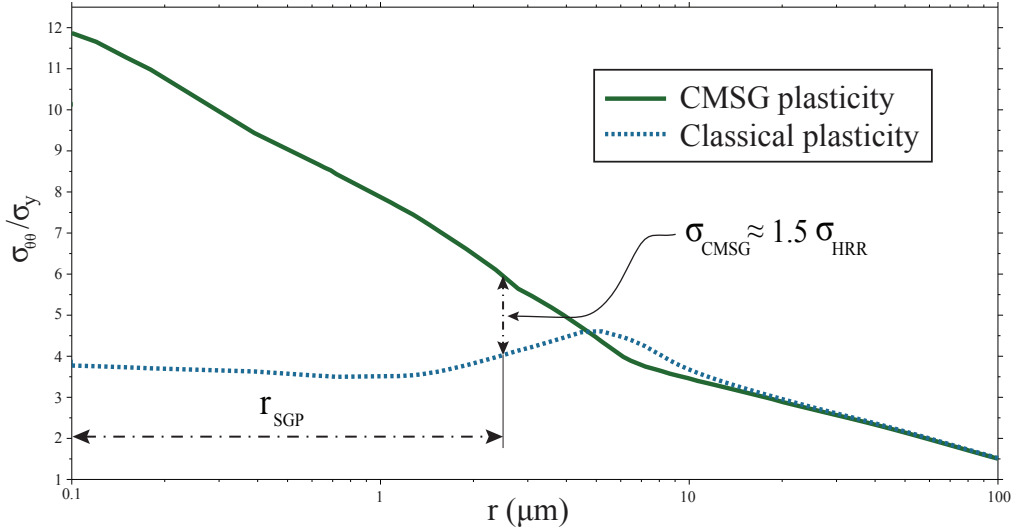}
\caption{$\sigma_{\theta\theta}$ distribution ahead of the crack tip for both CMSG and classical plasticity theories in large deformations, $r$ being the distance to the crack tip in log scale for $K_{I}=17.3\sigma_Y \sqrt{l}$, $\sigma_Y=0.2\%$ of $E$, $\nu=0.3$, $N=0.2$, and $l=3.53$ $\mu$m}
\label{fig:Fig5}
\end{figure}

To quantify the domain of influence of the strain gradient under large deformations, a parametric study is conducted. Furthermore, with the aim of establishing a comparison, results are obtained in the framework of the infinitesimal deformation theory; mimicking material properties and loading conditions. The variation of the normalized distance over which the strain gradient significantly influences the stress distribution-as a function of material properties, constraint conditions, and the applied load-is plotted in Fig. 6. Following the works by \citet{M97}, a relation between the crack tip and outer radii ($R/r=10^5$) is considered and a sufficiently higher upper bound for the load range ($K_{I}=1.2\sigma_Y^* \sqrt{R}$) is chosen to ensure a final blunting five times larger than the initial radius. Since the same range of values used for each parameter in section 3 is also considered in this case, results obtained can be compared with those shown in Fig. 4.

\begin{figure}
\makebox[\linewidth][c]{%
        \begin{subfigure}[b]{0.55\textwidth}
                \centering
                \includegraphics{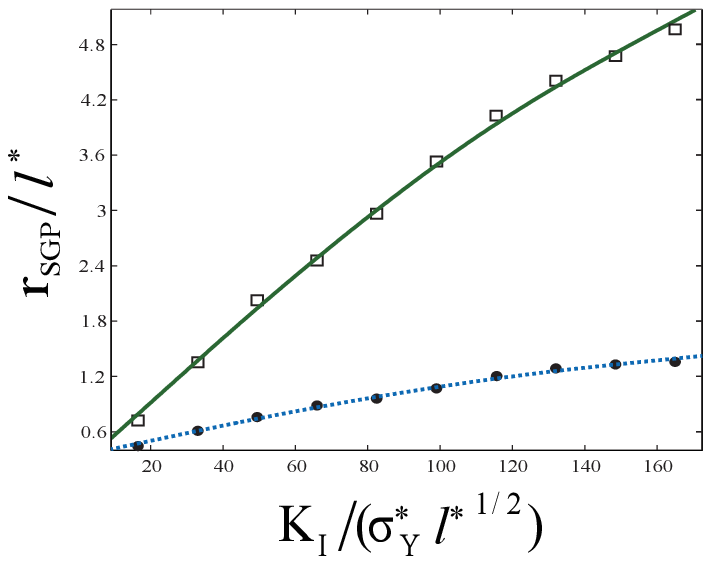}
                \caption{}
                \label{fig:Fig6a}
        \end{subfigure}
        \begin{subfigure}[b]{0.55\textwidth}
                \centering
                \includegraphics{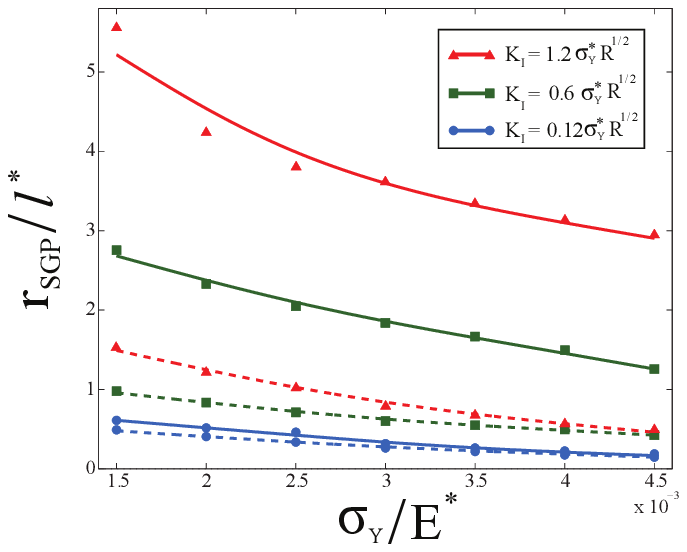}
                \caption{}
                \label{fig:Fig6b}
        \end{subfigure}%
       }
       
       \makebox[\linewidth][c]{%
        \begin{subfigure}[b]{0.55\textwidth}
                \centering
                \includegraphics{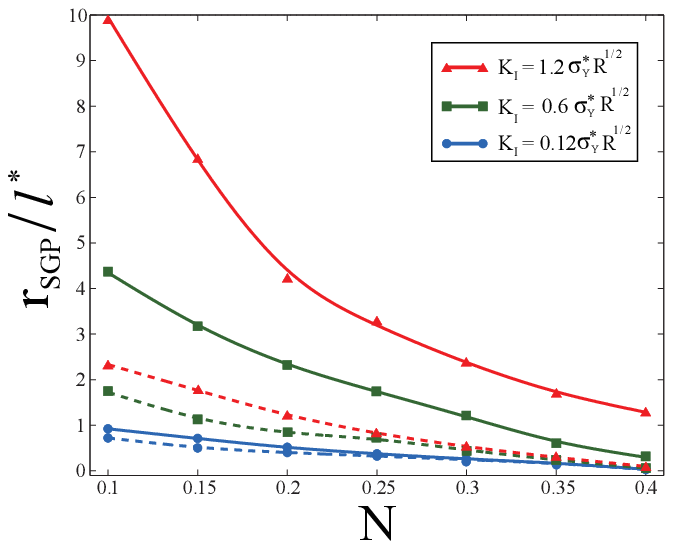}
                \caption{}
                \label{fig:Fig6c}
        \end{subfigure}
        \begin{subfigure}[b]{0.55\textwidth}
                \centering
                \includegraphics{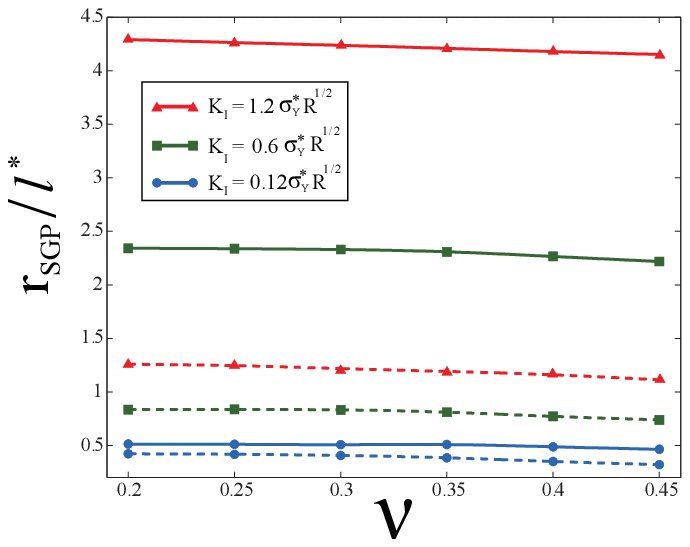}
                \caption{}
                \label{fig:Fig6d}
        \end{subfigure}%
        }
        
        \makebox[\linewidth][c]{%
        \begin{subfigure}[b]{0.55\textwidth}
                \centering
                \includegraphics{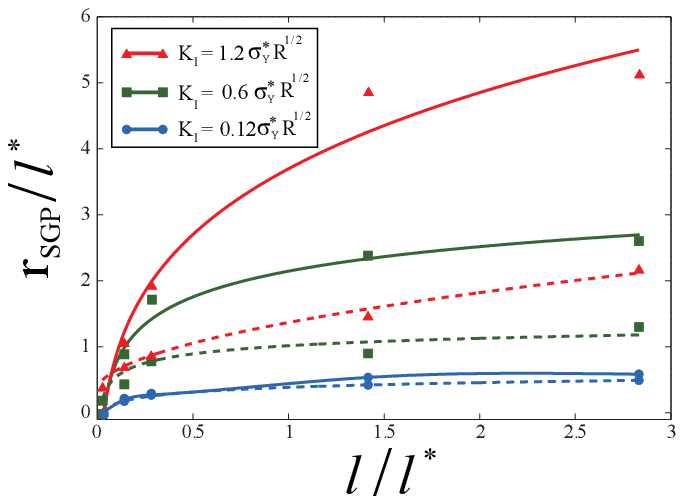}
                \caption{}
                \label{fig:Fig6e}
        \end{subfigure}
        \begin{subfigure}[b]{0.55\textwidth}
                \centering
                \includegraphics{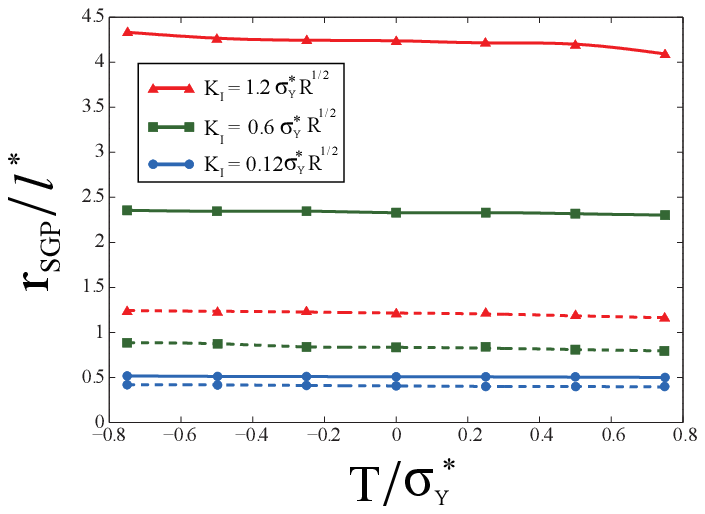}
                \caption{}
                \label{fig:Fig6f}
        \end{subfigure}%
        }

        \caption{Distance ahead of the crack tip where the strain gradient significantly influences the stress distribution under small (dashed lines) and large (solid lines) strains as a function of (a) applied load $K_I$, (b) yield stress $\sigma_Y$, (c) strain hardening exponent $N$, (d) Poisson's ratio $\nu$, (e) intrinsic material length $l$, and (f) $T$-stress. The material properties considered in Fig. 2 are considered as the reference values $(*)$.}\label{fig:Fig6}
\end{figure}

The trends shown in Fig. 6 for both large (solid lines) and small (dashed lines) strains are the same as those obtained for the parametric analysis in section 3. However, significantly higher values of $r_{SGP}$ are obtained in all cases when large deformations are considered. These results reveal that even in the case of a very small load ($K_{I}=0.12\sigma_Y^* \sqrt{R}$), accounting for large strains brings a relatively meaningful influence of the strain gradient. These differences with respect to the predictions that could be expected from the classical plasticity theory are much higher for load levels relevant to fracture and damage in metals ($K_{I}=0.6\sigma_Y^* \sqrt{R}$ and $K_{I}=1.2\sigma_Y^* \sqrt{R}$). Moreover, the results show a high sensitivity of the plastic size effect to the material properties and the applied load, so that a parametric study within the finite deformation theory is essential to rationally assess the need to incorporate an intrinsic material length in the continuum analysis.

\section{Discussion}
\label{Results implications}

The parametric study shows that higher values of the applied load and the intrinsic material length increase the influence of the strain gradients on crack-tip fields, whereas the opposite is true for the yield stress; the strain hardening exponent and the Poisson's ratio, being $r_{SGP}$ less sensitive to the latter parameter. Results concerning the yield stress are especially relevant since the hydrostatic stress follows the same trends. Therefore, the plastic size effect could strongly influence the process of hydrogen embrittlement, which severely degrades the fracture resistance of high strength steels. This is due to the central role that the stress field close to the crack tip plays on both hydrogen concentration and interface decohesion \citep{LG07}. Also, while results obtained within the infinitesimal deformation theory show that the effect of plastic strain gradient is negligible for higher values of $\sigma_Y$, which are common to high strength metallic alloys, strong differences arise between the stress fields of SGP and conventional plasticity theories when large strains are considered. This demonstrates the need to include the plastic size effect in the modelization of hydrogen-assisted cracking in metals. It is important to note that hydrogen-assisted damage occurs very close to the crack tip, the critical distance being lower than 1 $\mu$m (see e.g. \citealp{G03}) where the magnitude of stress elevation due to the influence of the strain gradient is significant. The ratio between the CMSG and the classical plasticity predictions is plotted in Fig. 7, where one can also notice that much higher values are obtained when large strains are considered.

\begin{figure}[h]
\makebox[\linewidth][c]{%
        \begin{subfigure}[b]{0.55\textwidth}
                \centering
                \includegraphics{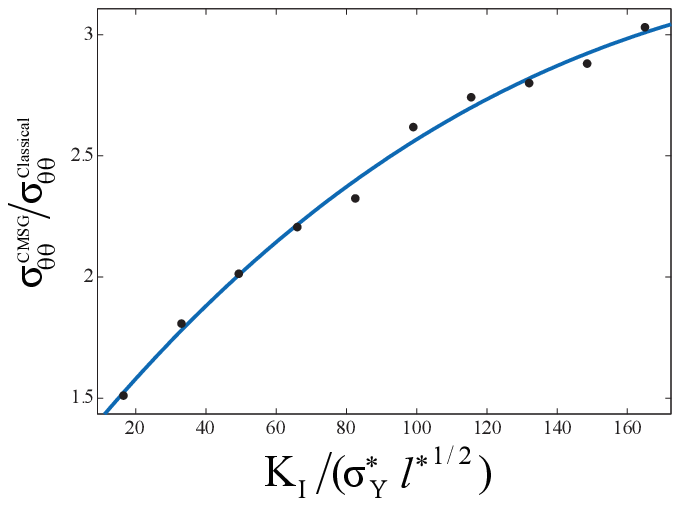}
                \caption{}
                \label{fig:Fig7a}
        \end{subfigure}
        \begin{subfigure}[b]{0.55\textwidth}
                \centering
                \includegraphics{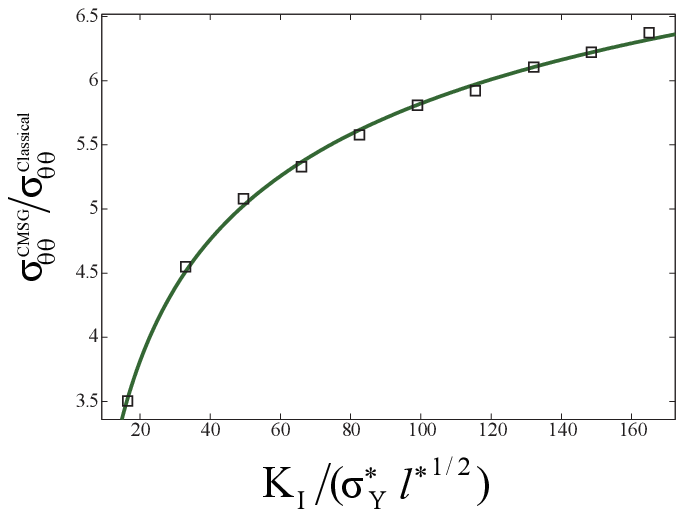}
                \caption{}
                \label{fig:Fig7b}
        \end{subfigure}
        }
       
        \caption{Ratio between the $\sigma_{\theta\theta}$ predictions of CMSG and classical plasticity at $r=0.1 \mu$m ahead of the crack tip ($\theta=0$) as a function of the applied load for (a) small strains and sharp crack and (b) large strains and blunted crack. The material properties are the same as those in Figure 2}\label{fig:Fig7}
\end{figure}

Previous works established that the domain where SGP effects can significantly elevate stresses over the HRR result for small strains was confined to distances less than 10 $\mu$m from the crack tip \citep{K08}. However, results shown in section 4 reveal that, when finite strains are considered, stress elevations persist to distances that could be one order of magnitude higher than those obtained within the infinitesimal deformation theory. This could have important implications on fracture and damage modeling of metals since the area where the strain gradient would significantly alter the crack-tip fields could span several voids ahead of the crack, and therefore influence various damage mechanisms that are characteristic of ductile fracture. Thus, results obtained from this work reveal that in the presence of a crack, near-tip stress-elevation that are predicted by SGP theories could significantly influence the probability of cleavage fracture in ductile-to-brittle transition analyses \citep{B08}, the prediction of stress-controlled nucleation of voids at large inclusions \citep{CN80}, the value of the parameters intrinsic to micromechanical failure models \citep{G75,TN84} when fitted through a top-down approach, or the onset of damage in stress-related coalescence criteria \citep{T90}. This is unlike previous studies on cleavage fracture and void growth, which did not consider the influence of the plastic strain gradient in modelization. 

Results concerning the MBL formulation (Figs. 3f and 6f) reveal that the aforementioned influence of the strain gradient on crack-tip fields remains under different constraint conditions since the size of the domain where significant differences between the stress fields of the SGP and the conventional plasticity theories arise is almost insensitive to changes in the $T$-stress value.

\section{Conclusions}
\label{Concluding remarks}

In this work, the influence of SGP theories on the fracture process of metallic materials has been numerically analyzed for both small and large deformations. The extensive parametric study conducted relates material properties, constraint scenarios, and applied loads with the physical distance ahead of the crack tip where the strain gradient significantly influences the stress distribution, thus identifying the conditions where the plastic size effect should be included in crack-tip damage modeling. 

Moreover, the incorporation of large strains and finite geometry changes in the numerical model reveals a meaningful increase in the domain influenced by the size effect, which may indicate the need to take consider the influence of the plastic strain gradient in the modelization of damage mechanisms, which has not been considered so far in the literature. 

\section{Acknowledgments}
\label{Acknowledge of funding}

The authors gratefully acknowledge the financial support from the Ministry of Science and Innovation of Spain through grant MAT2011-28796-CO3-03. The first author also acknowledges financial support from the University of Oviedo (UNOV-13-PF). 





\end{document}